\documentclass[twocolumn,showpacs,prl]{revtex4}

\usepackage{graphicx}

\begin{document}

\preprint{APS/123-QED}

\title{Full Aging in Spin Glasses}

\author{G.~F.~Rodriguez}

\author{G.~G.~Kenning}
\affiliation{Department of Physics, University of California\\
Riverside, California 92521-0101}
\author{R.~Orbach}

\affiliation{Department of Energy, Office of the Director of Science,\\
Washington DC\\}

\date{\today}

\begin{abstract}
The discovery of dynamic memory effects in the magnetization
decays of spin glasses in 1983 marked a turning point in the study
of the highly disordered spin glass state. Detailed studies of the
memory effects have led to much progress in understanding the
qualitative features of the phase space. Even so, the exact nature
of the magnetization decay functions have remained elusive,
causing confusion. In this letter, we report strong evidence that
the Thermoremanent Magnetization (TRM) decays scale with the
waiting time, $t_{w}$. By employing a series of cooling protocols,
we demonstrate that the rate at which the sample is cooled to the
measuring temperature plays a major role in the determination of
scaling.  As the effective cooling time, $t_{c}^{eff}$, decreases,
$\frac {t}{t_{w}}$ scaling improves and for $t_{c}^{eff}<20s$ we
find almost perfect $\frac{t}{t_{w}}$ scaling, i.e full aging.
\end{abstract}

\maketitle

Since the discovery of aging effects in spin glasses approximately
twenty years ago\cite{Cham83}\cite{Lund83}, much effort has gone
into determining the exact time dependence of the memory decay
functions. In particular, memory effects show up in the
Thermoremanent Magnetization (TRM) (or complementary Zero-Field
Cooled (ZFC) magnetization) where the sample is cooled through its
spin glass transition temperature in a small magnetic field (zero
field) and held in that particular field and temperature
configuration for a waiting time, $t_{w}$. At time $t_{w}$, a
change in the magnetic field produces a very long time decay in
the magnetization. The decay is dependent on the waiting time.
Hence, the system has a memory of the time it spent in the
magnetic field. A rather persuasive argument\cite{Bou92} suggests
that for systems with infinite equilibration times, the decays
must scale with the only relevant time scale in the experiment,
$t_{w}$.  This would imply that plotting the magnetization on a
t/$t_{w}$ axis would collapse the different waiting time curves
onto each other.  This effect has not been observed.

What has been observed\cite{Alba86} is that the experimentally
determined magnetization decays will scale with a modified waiting
time, $(t_{w})^\mu$. Where $\mu$ is a fitting parameter.  For
$\mu<1$ the system is said to have subaged. A $\mu>1$ is called
superaging and $\mu=1$ corresponds to full aging. For TRM
experiments a $\mu$ of approximately 0.9 is found for different
types of spin glasses\cite{Alba86, Ocio85}, over a wide range of
reduced temperatures indicating subaging. At very low temperatures
and temperatures approaching the transition temperature, $\mu$ is
observed to decrease from the usual 0.9 value. Superaging has been
observed in Monte Carlo simulations of spin glasses\cite{Sibani}.
This has led to confusion as to the exact nature of scaling.

Zotev et al.\cite{Zov02} have suggested that the departures from
full $\frac{t}{t_{w}}$ scaling, observed in aging experiments, are
mainly due to cooling effects. In a real experimental environment,
the situation is complicated by the time it takes for the sample
to cool to its measuring temperature. An effect due to the cooling
rate at which the sample temperature approaches the measuring
temperature has been known\cite{Nord87,Nord00}.  This effect is
not trivial, it does not contribute a constant time to $t_{w}$.

Another possible explanation for the deviation from full aging
comes from the widely held belief that the magnetization decay is
an additive combination of a stationary term ($M_{Stat} = A
(\tau_{0} / t)^{\alpha})$ and an aging term ($M =
f(\frac{t}{t_{w}})$~)\cite{Cou95,Bou95,Vin96}. Subtraction of a
stationary term, where $\tau_{0}$ is a microscopic spin flipping
time, A is a dimensionless constant and $\alpha$ is a parameter
determined from $\chi''$ measurements, was shown to increase $\mu$
from 0.9 to 0.97\cite{Vin96}.

In this letter we analyze effects of the cooling time through a
series of different cooling protocols and we present the first
clear and unambiguous experimental evidence that the TRM decays
scale as $\frac{t}{t_{w}}$(i.e. full aging).

\begin{figure*}
\scalebox{.58}{\includegraphics{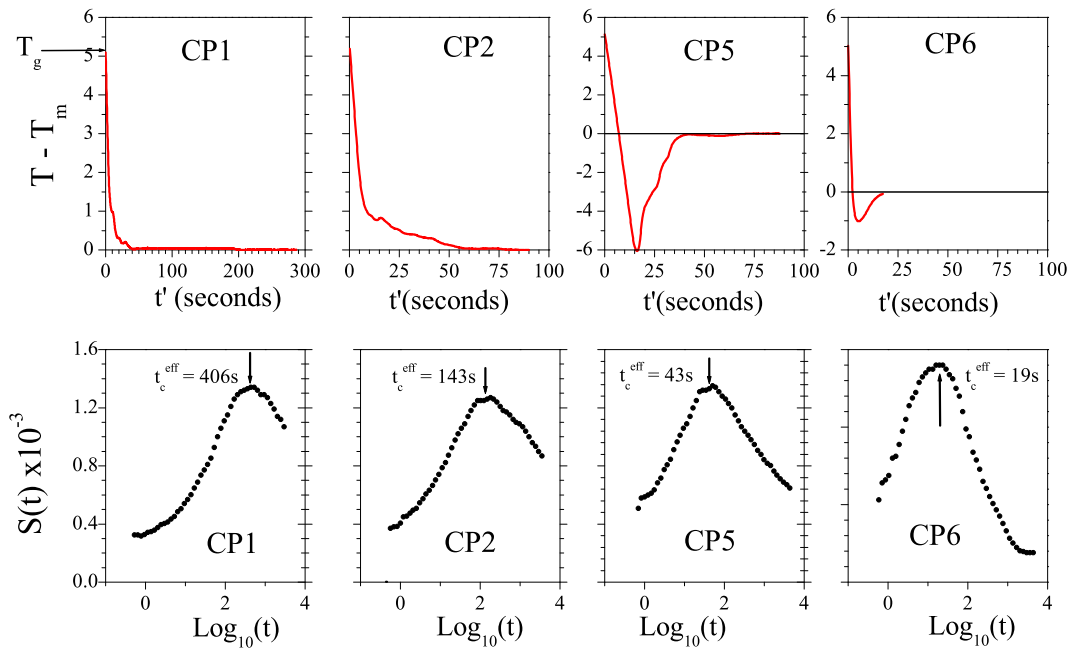}}
\caption{\label{fig:Cooling} The cooling time(t') start when
$T_{g}$ is crossed. For CP1-2 the temperature is never dropped
below $T_{m}$.  On the contrary for CP5-6 the temperature is not
allowed to go above $T_{m}$. The effective cooling time,
$t_{c}^{eff}$ is associated with a peak in the ZTRM S(t). The S(t)
for each each cooling protocol are plotted with the the same
y-scale.}
\end{figure*}

Three different methods have been regularly employed to understand
the scaling of the TRM decays. The first and simplest is to scale
the time axis of the magnetization decay with the time the sample
has spent in a magnetic field (i.e.$\frac{t}{t_{w}}$)\cite{Bou92}.
If the decays scale as a function of waiting time it would be
expected that the decay curves would overlap. This has not yet
been observed.

A second more sophisticated method was initially developed by
Struik\cite{Stu79} for scaling the dynamic mechanical response in
glassy polymers and first applied to spin glasses by Ocio et
al.\cite{Ocio85}. This method plots the log of the reduced
magnetization $M/M_{fc}$ ($M_{fc}$ is the field cooled
magnetization), against an effective waiting time
$\xi=\frac{\lambda}{t_{w}^{\mu}}$ where

\begin{eqnarray}
\lambda = \frac{t_{w}}{1-\mu}[(1+\frac{t}{t_{w}})^{1-\mu} -
1];~~~~\mu < 1 \label{eq:one}
\end{eqnarray}
or
\begin{eqnarray}
\lambda = t_{w} \log[1+\frac{t}{t_{w}}];~~~~~~~~~~~~~~~~\mu = 1
\label{eq:two}
\end{eqnarray}
A value of $\mu$ = 1 would correspond to perfect $t/t_{w}$
scaling. Previous values of $\mu$ obtained on the decays using
this method have varied from .7 to .94\cite{Alba86} for a
temperature range $.2<T_{r}<.95$. A value of $\mu<1$ is called
subaging.

Finally a peak in the function S(t)

\begin{eqnarray}
S(t)=-\frac{1}{H}\frac{dM(t)}{d[Log_{10}(t)]} \label{eq:three}
\end{eqnarray}

as a function of time has been shown to be an approximately linear
function of the waiting time\cite{Lund83}. This peak occurs at a
time slightly larger then the waiting time, again suggesting
possible subaging.

In this study we use all three of the above scaling procedures to
analyze the data we have produced with different cooling
protocols. All measurements in this letter were performed on our
homebuilt DC SQUID magnetometer with a $Cu_{.94}Mn_{.06}$ sample.
The sample is well documented\cite{Ken91} and has been used in
many other studies. The measurements described in this letter were
performed in the following manner: The sample was cooled, in a
magnetic field of 20 G, from 35 K through its transition
temperature of 31.5 K to a measuring temperature of 26 K. This
corresponds to reduced temperature of .83 $T_{g}$. The sample was
held at this temperature for a waiting time $t_{w}$, after which
time the magnetic field was rapidly decreased to 0G. The resulting
magnetization decay is measured 1s after field cutoff to a time
greater than or equal to 5$t_{w}$. The only parameters we have
varied in this study are $t_{w}$ and the rate and profile at which
we cool the sample through the transition temperature to the
measuring temperature. The sample is located on the end of the
sapphire rod and sits in the upper coil of a second order
gradiometer configuration. The temperature measuring thermometer
is located 12.5 cm above the sample. Heat is applied to the sample
through a heater coil located on the same sapphire rod 17 cm above
the sample. Sample cooling occurs by heat transfer with the He
bath via a constant amount of He exchange gas, which was
previously introduced into each chamber of the double vacuum
jacket. We have measured the decay time of our field coil and find
that we can quench the field in less then 0.1ms. We have also
determined that without a sample, our system has a small
reproducible exponential decay, that decays to a constant value
less than the system noise within 400 seconds. In order to
accurately describe our data we subtract this system decay from
all of the data.

\begin{figure}
\scalebox{.39}{\includegraphics{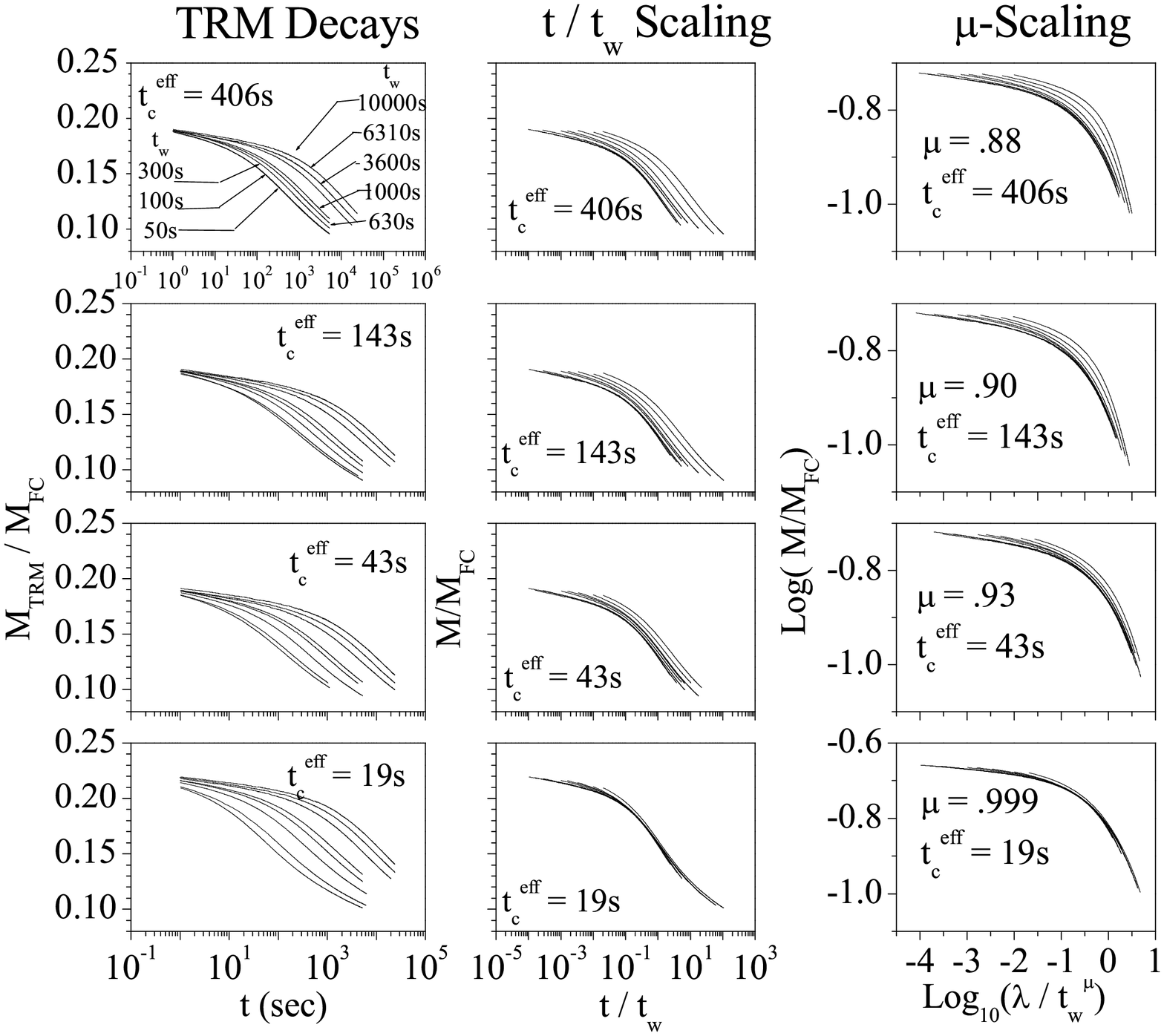}} \caption{\label{fig:Two} In
the first column the TRM decays curves for $t_{w}$ = 50s, 100s,
300s, 1000s, 3600s, 6310s, 10000s are plotted for the different
cooling protocols. In the middle column the decays are scaled with
the waiting. The y-axis for columns one and two are the same. The
decays collapse onto each other as $t_{c}^{eff}$ decreases. In the
third column the decays are scaled with $\mu$. Where $\lambda$
from eq~\ref{eq:one} one is used.}
\end{figure}

In this paper we present TRM data for eight waiting times
($t_{w}$= 50s, 100s, 300s, 630s, 1000s, 3600s, 6310s, and 10000s).
The same TRM experiments were performed for six different cooling
protocols. In this paper we use four of the cooling protocols.
Figure 1 (top row) is a plot of temperature vs. time for four of
the cooling protocols. These different cooling protocols, were
achieved, by varying applications of heat and by varying the
amount of exchange gas in the vacuum jackets. A more detailed
description of the cooling protocols will be given in a followup
publication. In Figure 1 (bottom row) we plot S(t)
(Eq~$\ref{eq:three}.)$ of the ZTRM protocol (i.e. zero waiting
time TRM) in order to characterize a time associated with the
cooling protocol, $t_{c}^{eff}$. As observed in Figure 1 we have
achieved effective cooling times ranging from 406s down to 19
seconds. These times can be compared with commercial magnetometers
which have cooling times in the range of 100-400s.

In Figure 2, we plot the data for the TRM decays (first column)
with the four cooling protocols. It should be noted that the
magnetization (y-axis) is scaled by the field cooled
magnetization.

The second column is the same data, as column one, with the time
axis(x-axis) normalized by $t_{w}$. It can be observed for
$\frac{t}{t_{w}}$ scaling (column 2) that as the effective cooling
time decreases the spread in the decays decreases giving almost
perfect $\frac{t}{t_{w}}$ scaling for the 19 second cooling
protocol.

The last column in Figure 2, is the data scaled,using $\mu$
scaling which has previously been described. It has long been
known that the rate of cooling affected $\mu$ scaling and that
$\mu$ scaling is only valid in the limit $t_{w}>>t_{c}^{eff}$. We
find this to be true and that the limit is much more rigorous than
previously believed.  To determine $\mu$ scaling, we focused on
applying this scaling to the longest waiting time data (i.e.,
$t_{w}$ = 3600s, 6310s and 10,000s). For the largest effective
cooling time data, $t_{c}^{eff}=406s$, we find that we can fit the
longest waiting time data with a $\mu$ value of .88. This is
consistent with previously reported values of $\mu$\cite{Alba86}.
We do find, however, that TRM data with waiting times less then
3600s do not fit on the scaling curve.  We find that scaling of
the three longest waiting time decays produces $\mu$ values which
increase as $t_{c}^{eff}$ decreases. We also find that as
$t_{c}^{eff}$ decreases, the data with shorter $t_{w}$ begins to
fit to the scaling better. It can be observed that at
$t_{c}^{eff}$=~19s we obtain almost perfect scaling for all of the
data with a value of $\mu=.999$. However we find we can reasonably
fit the data to a range of $\mu$ between .989-1.001. The fitting
for the large $t_{w}$ decay curves, $t_{w}$ = 3600s, 6310s and
10,000s, is very very good. Small systematic deviations, as a
function of $t_{w}$, occur for $t_{w}<3600s$ with the largest
deviations for $t_{w}$=~50s. Even with an effective cooling time
two orders of magnitude less then the waiting time, one sees
deviations from perfect scaling. We have also scaled the data
using Eq. 2. We find no noticeable difference between the quality
of this fit and the quality of the $\mu=.999$, for
$t_{c}^{eff}=19s$, shown in Figure 2. Data with longer cooling
times cannot be fit with Eq. 2. We therefore conclude that full
aging is observed for the long $t_{w}$ data using the
$t_{c}^{eff}=19s$ protocol.

It is clear from Figure 1 (bottom row), that the effect of the
cooling time has implications for the decay all the way up to the
longest time measured, 10,000 seconds. The form of the S(t) of the
ZTRM is very broad.  The S(t) function is often thought of as
corresponding to a distribution of time scales (or barrier
heights) within a system which has infinite equilibration times
(barriers). The peak in S(t) is generally associated with the time
scale (barrier height) probed in time $t_{w}$. In Figure 1, we
observe that for the larger effective cooling times, the waiting
times correspond to points on or near the peak of S(t) for the
ZTRM. We therefore believe that for the larger effective cooling
times there is significant contamination from the cooling protocol
over the entire region of $t_{w}$s used in this paper. Only for
$t_{c}^{eff}=19s$ cooling protocol do we find that the majority of
$t_{w}s$ occur far away from the peak in the S(t).

\begin{figure}
\resizebox{\columnwidth}{3.9in}{\includegraphics{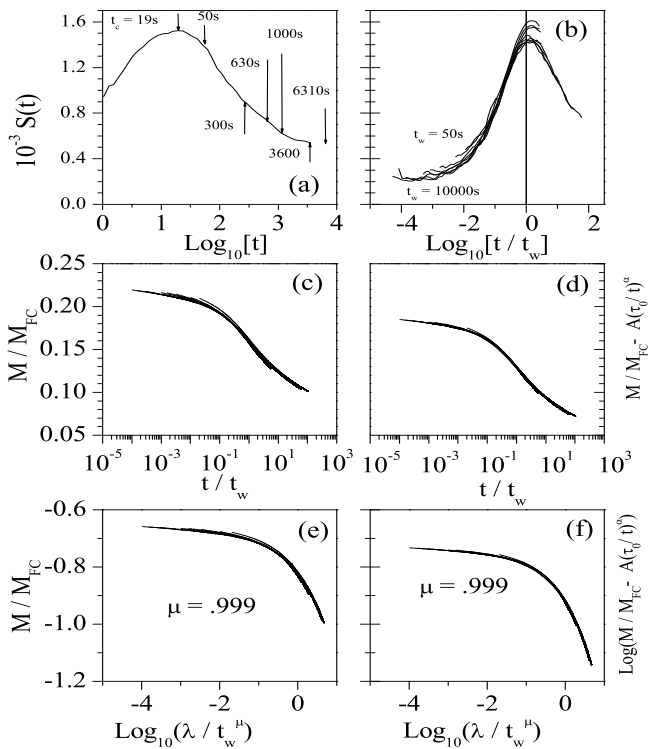}}

\caption{\label{fig:three} The S(t)(ZTRM) for $t_{c}$ =~19s is
shown with the different $t_{w}$s marked(a). In figure b. the S(t)
for all decays($t_{c}^{eff} =~19s$) are scaled with $t_{w}$. The
decay curves are scaled with $t_{w}$(c) and $\mu$-Scaled(e),
compared to the same scaling for decays with a stationary part
subtracted(d and f). For stationary part we use A = 0.06, $\alpha$
= 0.02 and $\tau_{0}=10^{-12}$}
\end{figure}
All the data in figure 3 used the cooling protocol with
$t_{c}^{eff} = 19s$. In Figure 3a we plot the S(t)(ZTRM) for
$t_{c}^{eff}=19s$ with arrows to indicate the waiting times for
the TRM measurements. It can be observed that after approximately
1000 seconds the slope of the S(t) function decreases, possibly
approaching a horizontal curve, which would correspond to a pure
logarithmic decay in M(t). If, on the other hand, the slope is
continuously changing this part of the decay may be described by a
weak power law. Either way, this region would correspond to aging
within a pure non-equilibrated state. We believe that the long
waiting time data occurs outside the time regime that has been
corrupted by the cooling time and that this is the reason that we
have, for the first time, observed full aging.

It has been suggested that subtraction of a stationary component
of the magnetization decay will improve scaling\cite{Vin96}. The
very long time magnetization decay is believed to consist of a
stationary term that is thought to decay as a power law. We fit a
power law, $M(t)=A(\tau_o/t)^\alpha$ to the long time decay
(1000s-5000s), of the ZTRM for $t_{w}$=19s. Using
$\tau_{o}=10^{-12}s$, we find $\alpha=.07$ and $A=.27$.
Subtracting this power law form from the magnetization decay
destroys scaling. We find that the subtraction of a much smaller
power law term with A=.06 and $\alpha$=.02 slightly improves
scaling at both short and long times. While the $\alpha$ values
for the two different power law terms we have fit to are quite
different, both values fall within the range determined from the
decay of $\chi$"\cite{Vin96}. In Figure 3(c-d) and 3(e-f), we plot
the two different types of scaling we have performed, with and
without the subtraction of the weaker power law term.

We find that even for $t_{c}^{eff}=19s$ the peak in S(t) for
$t_{w}>1000s$ occur at a time larger then $t_{w}$(fig 3b). We find
that we can fit the effective time associated with the peak in
S(t) to $t_{w}^{eff}=t_w^{1.1}$.

In summary, we have performed TRM decays over a wide range of
waiting times (50s~-~10,000s) for six different cooling protocols.
We find that as the time associated with the cooling time
decreases, scaling of the TRM curves improves in the
$\frac{t}{t_{w}}$ scaling regime and in the $\mu$ scaling regime.
In $\mu$ scaling we find that as the effective cooling time
decreases $\mu$ increases approaching a value of .999 for
$t_{c}^{eff}=19s$. For the $t_{c}^{eff}=19s$ TRM decays, we find
that subtraction of a small power law term (A=.06, $\alpha$=.02)
slightly improves the scaling. It is however likely that the small
systematic deviations of the $t_{c}^{eff}=19s$ data as a function
of $t_{w}$ are associated with the small but finite cooling rate.

The authors would like to thank V.~S.~Zotev, E.~Vincent and
J.~M.~Hammann for very helpful discussions.

\end{document}